\documentclass[11pt,fleqn]{article}
\usepackage{amssymb}
\usepackage[latin1]{inputenc}
\usepackage{times}
\usepackage{mathptm}
\newcommand{\dagg}{{\scriptscriptstyle\dagger}}
\begin{document}
\mathindent 0mm
%%%%%%%%%%%%%%%%%%%%%%%%%%%%%%%%%%%%%%%%
\title{\begin{flushright} \normalsize HD--TVP--98-11 \end{flushright}
  \vspace*{0.5cm} Ferromagnetism in single-band Hubbard models with a
  partially flat band} \author{Andreas Mielke\thanks{E--mail:
    mielke@tphys.uni-heidelberg.de} \\
  Institut f\"ur Theoretische Physik, Ruprecht--Karls--Universit\"at, \\
  Philosophenweg 19, D-69120~Heidelberg, F.R.~Germany} 
  \date{March 10, 1999}
\maketitle
\begin{center}
  Accepted for publication in Physical Review Letters
\end{center}
\begin{abstract}
  A Hubbard model with a single, partially flat band has ferromagnetic
  ground states. It is shown that local stability of ferromagnetism
  implies its global stability in such a model: The model has only
  ferromagnetic ground states if there are no single spin-flip ground
  states.  Since a single-band Hubbard model away from half filling
  describes a metal, this result may open a route to metallic
  ferromagnetism in single band Hubbard models.
  \\[0.5cm]
  PACS-numbers: 75.10.Lp, 71.10.Fd
\end{abstract}

Ferromagnetism of itinerant electrons is an old and still unsolved
problem in theoretical physics. One of the motivations to introduce
the Hubbard model \cite{Hubbard} more than 30 years ago has been to
understand this problem \cite{Kanamori}. Conceptually, the Hubbard
model is very simple.  It describes electrons on a lattice interacting
via a repulsive, purely local Coulomb interaction.  Originally, only a
single energy band was considered.  But it is relatively easy to
generalize the model including more energy bands or further
interactions.  Despite the simplicity of the model, there are only few
rigorous results concerning the existence of ferromagnetism. The first
rigorous result is the so called Nagaoka theorem \cite{Nagaoka}.  It
states that on certain lattices, for an infinitely large Coulomb
repulsion, and close to half filling, the ground state is
ferromagnetic. But the problem is that this ferromagnetic state
disappears if the repulsion has realistic values or if the number of
electrons is varied \cite{Suto,Variation}, at least for usual cubic
lattices in $d$ dimensions.\cite{LM}

Ferromagnetic ground states also occur, if one of several bands of the
Hubbard model is dispersionless (Lieb's ferrimagnetism \cite{Lieb} and
the flat-band ferromagnetism \cite{flat}).  Tasaki \cite{Tasaki} was
able to show the local stability of ferromagnetic ground states in
related models with a nearly flat band. But these models (as well as
the flat-band ferromagnetism) are already extensions of the single
band Hubbard model since they contain more than one band.
Furthermore, these models show characteristic features of an
insulator: The (nearly) flat band is half filled and
there is a gap in the single-particle spectrum. 
Other extensions of the Hubbard model have been
proposed as well, for instance an additional ferromagnetic interaction
between the electrons \cite{Wahle} or degenerate bands with Hund's
coupling between the bands \cite{Kubo}.

The question, if and under which conditions a simple, \emph{single}
band Hubbard model can show ferromagnetism, is still open.
Ferromagnetism is a strong coupling phenomenon. 
To obtain ferromagnetism within usual mean
field theory the dimensionless parameter $\rho_FU$ has to be large, 
where $\rho_F$ is the density of states at the
Fermi level.  But it should be noted that a large value of $\rho_FU$
is not sufficient for the existence of ferromagnetism (counterexamples
with $\rho_F=\infty$ and $U=\infty$ can be found in \cite{am_noferro}). 
On the other
hand it seems to be clear that a strong asymmetry of the band,
together with a large density of states near the Fermi energy is a
condition that favours ferromagnetism. This view is supported by the
variational calculations by Hanisch et al.  \cite{Variation} and as
well by the dynamical mean field analysis by Wahle et al. \cite{Wahle}.
In this letter I discuss the extreme limit of this situation, namely
a Hubbard model with a partially flat band. The main results
is that local stability of ferromagnetism implies its global
stability in such models.

At a first glance a
single, partially flat band seems to be 
similar to the flat-band ferromagnetism
mentioned above. But there are important differences: As already
mentioned, the models with a nearly flat band have a gap in the single
particle spectrum and the band is half filled \cite{Tasaki}. This is a
typical situation for an insulator. In the flat band case, it was
possible to go away from half filling \cite{flat}, but if an entire
band is flat, single particle eigenstates may be localized. In fact,
the existence of a localized basis was an essential point in the
proofs \cite{flat}.  Localized states are as well typical for an
insulator.  In the present case, the energy band is not half
filled and there is no gap in the single particle spectrum. A
localized eigen-basis of single particle states does not exist,
instead the translational invariance will be important. 
These are typical properties of a metal.

The Hamiltonian of a single band Hubbard model on a $d$-dimensional
translationally invariant lattice with periodic boundary conditions is
\begin{equation}
  H=\sum_{k\sigma}\epsilon_k c^{\dagg}_{k\sigma}c_{k\sigma}
  +\frac{U}{N_s}\sum_{\delta}
  \left(\sum_k c^{\dagg}_{k+\delta\uparrow}c^{\dagg}_{-k\downarrow}\right)
  \left(\sum_k c_{-k\downarrow}c_{k+\delta\uparrow}\right).
\end{equation}
$c^{\dagg}_{k\sigma}$ ($c_{k\sigma}$) creates (annihilates) an
electron with spin $\sigma$ in a single particle eigenstate given by
$\psi_k(x)=N_s^{-1/2}\exp(ikx)$, and $N_s$ is the number of sites of
the lattice, which is equal to the number of single particle states in
the band. The interaction is usually written in the form $U\sum_x
c^{\dagg}_{x\uparrow} c^{\dagg}_{x\downarrow} c_{x\downarrow}
c_{x\uparrow}$. It is repulsive, $U>0$.  The Hamiltonian has the
usual $SU(2)$ spin symmetry.  Furthermore, due to translational
invariance, the total momentum is a good quantum number.  The wave
vectors $k$ are elements of the first Brillouin zone (BZ).  More
precisely, $k$ is a representative of a class of equivalent wave
vectors each belonging to a different BZ. A statement like
$k+\delta=k'$ means that $k+\delta$ and $k'$ are both representatives
of the same class of wave vectors.  I assume that the single particle
band is partially flat, a finite fraction of the band
energies $\epsilon_k$ is degenerate.  I assume that the set of
degenerate band energies is situated at the bottom of the band, but
some comments on other situations are given as well.  Shifting the
energy scale one can always take $\epsilon_k\ge 0$.  As a consequence,
$H\ge 0$. Let $\cal{L}$ be the subset of wave vectors $k$ with
$\epsilon_k=0$. $N_d$ is the degeneracy, i.e. the number of elements
in $\cal{L}$. In such a situation the Hamiltonian has ferromagnetic
ground states if $N_e\le N_d$. Let
\begin{equation} \label{gs0}
  \psi_{0F}=\prod_{k \in \cal{L}}c^{\dagg}_{k\uparrow}|0\rangle
\end{equation}
be the (only) ground state of $H$ with $S=S_z=N_d/2$, $N_e=N_d$.  For
$N_e<N_d$ one can construct ${N_d \choose N_e}$ ferromagnetic ground
states with $S=S_z=N_e/2$ by replacing the product in (\ref{gs0})
by a product over an arbitrary subset of $\cal{L}$ with $N_e$
elements. Using the $SU(2)$ invariance of the
Hamiltonian one can construct further ground states with $S_z<S=N_e/2$. 
The question is: Are there ground states with $S<N_e/2$.

$\psi_{0F}$ has zero energy. Any other ground state with $N_e\le N_d$
must have zero energy as well. Since both parts of the Hamiltonian,
the kinetic energy and the interaction, are non-negative, the ground
states of $H$ are simultaneously ground states of the kinetic energy
and of the interaction. This simplifies the situation a lot.  Let us
assume that $H$ has a ground state with $N_e=N_d-n+m$ electrons ($n\ge
m$) and a spin $S=S_z=N_e/2-m$. Such a state can be written in the
form
\begin{equation}
  \label{ansatz}
  \psi=S_-^{n,m}(\alpha)\psi_{0F}
\end{equation}
where
\begin{equation}
  S_-^{n,m}(\alpha)=\sum_{l_1 \ldots l_m;k_1 \ldots k_n}
  \alpha_{l_1 \ldots l_m;k_1 \ldots k_n}
  \prod_jc^{\dagg}_{l_j\downarrow} \prod_j c_{k_j\uparrow}.
\end{equation}
$\alpha_{l_1 \ldots l_m;k_1 \ldots k_n}$ are antisymmetric in the
first $m$ and in the last $n$ indices.  $\psi$ should be a state with
spin $S=S_z=N_e/2-m$. This is the case if $S_+\psi=0$ where $S_+=\sum_k
c^{\dagg}_{k\uparrow} c_{k\downarrow}$, which yields $\sum_k
\alpha_{k,l_2 \ldots l_m;k,k_2 \ldots k_n}=0$.  Since the Hamiltonian is
translationally invariant, the eigenstates of $H$ are also eigenstates
of the momentum operator. Let $\psi$ be a state with
momentum $p$. This means that $\alpha_{l_1 \ldots l_m;k_1 \ldots k_n}$
vanishes if $\sum_{j=1}^{m}l_j-\sum_{j=1}^{n}k_j\ne p$. Since
$\psi$ is a ground state of the kinetic energy, 
$\alpha_{l_1 \ldots l_m;k_1 \ldots k_n}$ has to vanish if some indices
$l_j$ are not in $\cal{L}$. Furthermore, I let  
$\alpha_{l_1 \ldots l_m;k_1 \ldots k_n}=0$ if some indices
$k_j$ are not in $\cal{L}$. $\psi$ is a ground
state of the interaction if and only if $\sum_k c_{k+\delta\uparrow}
c_{-k\downarrow} \psi=0$ for all $\delta$.  This yields a condition
for $\alpha$, namely
\begin{equation}
  \label{cond1s}
  \sum_{P\in S_{n+1}}(-1)^P
  \alpha_{l_1 \ldots l_{m-1},-k_{P(n+1)}+\delta;k_{P(1)} \ldots k_{P(n)}}=0
\end{equation}
for all $k_j$, $l_j$, $\delta$. $S_{n+1}$ is the group of all
permutations $P$ of $n+1$ objects and $(-1)^P$ denotes the sign of the
permutation $P$; $(-1)^P=1$ if $P$ is even and $-1$ if $P$ is odd.
I let $\delta=\sum_{j=1}^{n+1}k_j-\sum_{j=1}^{m-1}l_j+p$, since otherwise
(\ref{cond1s}) is trivial. Using the fact that $\alpha$ is
antisymmetric in the last $n$ indices, one can rewrite (\ref{cond1s})
in the form
\begin{equation}
  \label{cond2s}
  \sum_{i=1}^{n+1}(-1)^{n(i-1)}\alpha_{l_1\ldots l_{m-1},
    p+\sum_{j=1}^{n+1}k_j-\sum_{j=1}^{m-1}l_j-k_i;
    k_{i+1}\ldots k_{n+1},k_1\ldots k_{i-1}}=0.
\end{equation}
The sum runs now over all cyclic permutations of the indices $k_j$. 
I define
\begin{equation}
  \tilde{\alpha}_{l_2 \ldots l_m;k_2 \ldots k_n}=
  \sum_{k_1}
  \alpha_{k_1+\tilde{p},l_2\ldots l_m;k_1 \ldots k_n}
\end{equation}
where $\tilde{p}\ne 0$ is chosen such that $\tilde{\alpha}$ is not
identically zero. 
This is possible since for some $k\in \cal{L}$,
$k+\tilde{p}$ is also in $\cal{L}$ and since $\alpha$ is not
identically zero \cite{tildealpha}. 
I put $l_1=k_1+\tilde{p}$ in (\ref{cond2s}) and sum
over $k_1$. Using the definition of $\tilde{\alpha}$ and the
antisymmetry of $\alpha$ in the last $n$ indices, one obtains
\begin{eqnarray}
  &\,&\sum_{k_1}
  \alpha_{k_1,l_2\ldots l_{m-1},
    p+\sum_{j=2}^{n+1}k_j-\sum_{j=2}^{m-1}l_j-k_1;
    k_2\ldots k_{n+1}}
  \nonumber \\ 
  &-&
  \sum_{i=2}^{n+1}(-1)^{(n-1)(i-2)}\tilde{\alpha}_{l_2\ldots l_{m-1},
    p-\tilde{p}+\sum_{j=2}^{n+1}k_j-\sum_{j=2}^{m-1}l_j-k_i;
    k_{i+1}\ldots k_{n+1},k_2\ldots k_{i-1}}=0.
\end{eqnarray}
The first term in this equation vanishes. The reason is that for each
$k_1$ there is a term in the sum over $k_1$ with $k_1$ replaced by
$p+\sum_{j=2}^{n+1}k_j-\sum_{j=2}^{n-1}l_j-k_1$. Due to the
antisymmetry of $\alpha$ in the first $m$ indices, these two terms
are equal up to a different sign and
annihilate each other. This yields
\begin{equation}
  \label{nton-1}
  \sum_{i=2}^{n+1}(-1)^{(n-1)(i-2)}\tilde{\alpha}_{l_2\ldots l_{m-1},
    p-\tilde{p}+\sum_{j=2}^{n+1}k_j-\sum_{j=2}^{m-1}l_j-k_i;
    k_{i+1}\ldots k_{n+1},k_2\ldots k_{i-1}}=0
\end{equation}
which is the same condition as (\ref{cond2s}) for $\tilde{\alpha}$
instead of $\alpha$. Consequently
$\tilde{\psi}=S_-^{n-1,m-1}(\tilde{\alpha})\psi_{0F}$ is also a ground
state of $H$. This shows that if $H$ has a ground state with
$N_e=N_d-n+m$ electrons ($n\ge m$) and a spin $S=S_z=N_e/2-m$, it has
also a ground state with the same number of electrons and a spin
$S=S_z=N_e/2-m+1$.  One can now iterate this procedure to obtain
finally a single spin flip state with a spin $S=S_z=N_e/2-1$:

{\it Theorem} -- In a single band Hubbard model with a $N_d$-fold
degenerate single particle ground state and $N_e\le N_d$ electrons 
local stability of ferromagnetism implies global stability. The model has
only ferromagnetic ground states with a spin $S=N_e/2$, if there are
no ground states with a single spin flip, i.e. with a spin
$S=N_e/2-1$.

{\it Remarks} -- 1. The existence of single spin flip ground states is
relatively easy if $N_e=N_d$.  A general single spin flip state with
momentum $p$ is given by
\begin{equation}
  \psi=\sum_k \alpha_{k+p,k} c^{\dagg}_{k+p\downarrow}
  c_{k\uparrow} \psi_{0F}.
\end{equation}
This state has a spin $S=N_e/2-1$ if $p\ne 0$.  The condition
(\ref{cond1s}) shows that this state is a ground state if and only if
$\alpha_{k+p,k}=\alpha_{k'+p,k'}$ for all $k,k'\in \cal{L}$.
Therefore, it is possible to construct a single spin flip ground state
with momentum $p$ if and only if $\epsilon_{k+p}=0$ for all $k$ with
$\epsilon_k=0$.  In that situation, the single particle density matrix
in \cite{AM1} is reducible. Thus, for $N_e=N_d$ our result is as well
a consequence of the result in \cite{AM1}, but due to translational
invariance, the condition for the occurrence of ferromagnetism is much
simpler.  For $N_e<N_d$ it is not possible to obtain a similar
(simple) condition for the existence of a single spin flip ground
state.

2. Let us consider situation where the degenerate single particle energy
lies at the upper band edge of the single band, and let $N_e\ge 2N_s-N_d$. 
In this case the
model has ferromagnetic ground states with a spin $S=(2N_s-N_e)/2$. 
Performing a particle--hole transformation one obtains a model that
fulfils the conditions of the above theorem. Thus local stability
of ferromagnetism implies global stability in this case as well.

3. In the above derivation one uses the fact, that the Hamiltonian is
translationally invariant. This is a natural assumption.  But it is
also possible to investigate a more general case.  As for $N_e=N_d$ in
\cite{AM1}, the proof is much more complicate and less intuitive
\cite{AMf}.

4. The result is true for any $U>0$. $U$ may be arbitrarily small.
Therefore one may wonder whether this model corresponds to a strong
coupling situation. This is indeed the case. The relevant
dimensionless parameter is $\rho_FU$. It is infinite in our model for
any $U>0$, since the Fermi level lies in the region where the band is
flat.  A partially flat band is certainly an unrealistic situation.
Typically, this assumption has the consequence that the hopping matrix
elements $t_{xy}=N_s^{-1}\sum_k\epsilon_k\exp(ik(x-y))$ have a longer
range than usual. On the other hand, one may hope that as in the flat
band case \cite{Tasaki} our result extends to a nearly flat case as
long as $U$ is not too small. This would be a realistic situation in
transition metals.

5. In most cases it is much simpler to study the stability of the
ferromagnetic state with respect to single spin flips than the global
stability. In the case of a Hubbard model with a nearly flat band,
Tasaki \cite{Tasaki} was able to show the stability with respect to
single spin flips. In most variational treatments single spin flip
states are used as well \cite{Variation}.  The variational studies of
the stability of the Nagaoka state \cite{Suto,Variation} are
complicated and the general single spin flip problem is too difficult
to be solved completely on usual lattices in $1<d<\infty$ dimensions.
In our case the situation is simpler since we already know that there
are ferromagnetic ground states.  The aim is only to show under which
conditions there are other ground states. If there are further ground
states, one should expect that small perturbations are sufficient to
destroy ferromagnetism. This would be an instable situation. If there
are no other ground states, it is possible that ferromagnetism is
stable with respect to small perturbations of the Hamiltonian. But it
is difficult to investigate this problem since in the present model
there is no gap in the single particle spectrum. This is the main
technical difference to the models discussed in \cite{Tasaki}.

6. As in \cite{AM1} one can generalize the above result to a
situation, where the degenerate single particle states are not at the
bottom of the band. I assume again that the $N_d$-fold degenerate
single particle state has energy 0 and that $\cal{L}$ is the subset of
$k$ with $\epsilon_k=0$. Let $\cal{L_<}$ be the subset of $k$ with
$\epsilon_k<0$ and let $N_<$ be the number of elements of $\cal{L_<}$.
For $U=0$ and $N_e\le 2N_<+N_d$ the ground states are highly
degenerate: Each single particle state with $k\in \cal{L_<}$ contains
two electrons and the remaining $N_e-2N_<$ electrons can be
distributed arbitrarily among the states with zero energy. If $U$ is
small one can apply degenerate perturbation theory.  This means that
among these degenerate states one has to find those with a minimal
interaction energy. Since the system is translationally invariant the
contribution from the single particle states with $k\in \cal{L_<}$ is
the same for all the degenerate multi particle states. It is therefore
sufficient to minimize the interaction energy of the electrons in
single particle states with $k\in \cal{L}$.  This is equivalent to the
above situation, where $\epsilon_k=0$ was the bottom of the band. If
the degeneracy is lifted at first oder in $U$, the ground state is
ferromagnetic with a spin $S=(N_e-2N_<)/2$.  Depending on $N_e$, the
spin varies between $0$ and $N_d/2$ and may be extensive if $N_d$ is
extensive. If the degeneracy is not lifted at first order, another
ground state with a smaller spin is usually favoured at higher order
in $U$. This argument explains e.g. the results presented in
\cite{KA}. But this argument is perturbative and holds only for (very)
small $U$.  Is it possible that this ferromagnetism disappears when
$U$ becomes larger? In the moment I am not able to answer this
question. But if the following conjecture is true, the ferromagnetism
is stable for any finite $U$.

{\it Conjecture} -- Let $E_{0S}$ be the smallest eigenenergy of $H$ in
the subspace of eigenstates with a spin $S$. Suppose that for some
$U=U_0$ the Hamiltonian has a degenerate ground state with a ground
state energy $E_{0S_1}=E_{0S_2}$ and $S_1\ne S_2$.  Then I claim that
$E_{0S_1}\le E_{0S_2}$ for $U>U_0$ if $S_1>S_2$.

It is sufficient to proof this conjecture for $N_e\le N_s$, since the
result for larger electron numbers can be obtained using a
particle-hole transformation.  I am not aware of any (rigorous) result
for the Hubbard model that is in contradiction to this conjecture.
The physical intuition behind it is simply that if for some value
$U_0$ of the interaction a degeneracy occurs, one should expect that
the state with a higher spin should have a larger kinetic energy and a
smaller interaction energy, so that for higher values of $U$ the
higher spin is favoured.  As far as I know there is no proof for this
conjecture.  The conjecture is trivial if $S_1=N_e/2$, since the
energy of a state with maximal spin (and $N_e\le N_s$) does not depend
on $U$, whereas any other eigenenergy is a monotonously increasing
function of $U$.

Let us note that this route to ferromagnetism in single band Hubbard
models naturally leads to non-saturated ferromagnetic states if the
degeneracy is not situated at a band edge. This is similar to Lieb's
ferrimagnetism \cite{Lieb}. It is even possible that a single band
Hubbard model on a bipartite lattice has a degeneracy somewhere in the
(symmetric) band.  But if this degeneracy occurs in the middle of the
band Lieb's theorem tells us that the ground state has $S=0$. This is
not a contradiction to the above result. In such a special situation
one can easily see that $p$ exists such that for each $k\in \cal{L}$,
$k+p$ is as well in $\cal{L}$. Therefore, as shown in the first
remark, the degeneracy is not lifted within a first order
perturbational treatment and a second order perturbational treatment
favours the singlet state, as predicted by Lieb's theorem.  The most
simple bipartite lattice, where this situation occurs, is a bipartite
lattice where $t_{xy}=t$ if $x$ and $y$ belong to different
sublattices and $t=0$ otherwise. Such a lattice is called a complete
bipartite graph.


\begin{thebibliography}{99}
\bibitem{LM} A special case is the one-dimensional Hubbard model, for
  which ferromagnetism is unstable in the whole parameter space.
\bibitem{tildealpha} The existence of $\tilde{\alpha}$ 
can be shown as follows: Define
$\tilde{\psi}_{\tilde{p}}=S_-^{n-1,m-1}(\tilde{\alpha})$
and calculate $\sum_{\tilde{p}}\|\tilde{\psi}_{\tilde{p}}\|^2$.
This quantity can be shown to be positive, so that at least
for some $\tilde{p}$ one has a non-trivial $\tilde{\alpha}$.
\end{thebibliography}
\end{document}